\documentclass[11pt,letterpaper]{article}
\usepackage[margin=1in]{geometry}
\usepackage[super,sort&compress,numbers]{natbib}

\usepackage[singlelinecheck=false]{caption}
\usepackage{setspace,array,verbatim,float,multirow}
\usepackage{ucs,amssymb,amsmath,mathtools}
\usepackage[utf8]{inputenc}
\usepackage{fontenc,graphicx}
\usepackage[usenames,pdftex,dvips]{color,xcolor}
\usepackage[pdftex,colorlinks]{hyperref,ragged2e}

\newcommand{\matr}[1]{\mathbf{#1}}
\newcommand{\Nb}{ \nobreak}

\newcommand{\Pv}{P\nobreakdash-value}

\newcommand{\Pvs}{P\nobreakdash-values}

\title{Detecting weak signals by combining small P-values in genetic association studies}
\author{Olga A. Vsevolozhskaya,$^1$ Fengjiao Hu,$^2$ Dmitri V. Zaykin$^{2\,\ast}$} 
\begin{document}

\doublespacing

\date{}
\maketitle
\thispagestyle{empty}

\noindent
$^1$Department of Biostatistics, College of Public Health, University of Kentucky, Lexington, KY 40536, USA. \\
\noindent
$^2$Biostatistics and Computational Biology, National Institute of Environmental Health Sciences, National Institutes of Health, Research Triangle Park, NC 27709, USA.\\
\vskip 2ex
\noindent
$^\ast$Correspondence: Dmitri V. Zaykin, Senior Investigator at the Biostatistics and Computational Biology Branch, National Institute of Environmental Health Sciences, National Institutes of Health, P.O. Box 12233, Research Triangle Park, NC 27709, USA. Tel.: +1 (919) 541-0096 ; Fax: +1  (919) 541-4311. Email address: dmitri.zaykin@nih.gov

\clearpage
\setcounter{page}{1}
\section*{Abstract}
We approach the problem of combining top-ranking association statistics or \Pvs{} from a new perspective which leads to a remarkably simple and powerful method.
Statistical methods, such as the Rank Truncated Product (RTP), have been developed for combining top-ranking associations and this general strategy proved to be useful in applications for detecting combined effects of multiple disease components. To increase power, these methods aggregate signals across top ranking SNPs, while adjusting for their total number assessed in a study. Analytic expressions for combined top statistics or \Pvs{} tend to be unwieldy, which complicates interpretation, practical implementation, and hinders further developments. Here, we propose the Augmented Rank Truncation (ART) method that retains main characteristics of the RTP but is substantially simpler to implement. ART leads to an efficient form of the adaptive algorithm, an approach where the number of top ranking SNPs is varied to optimize power. We illustrate our methods by strengthening previously reported associations of $\mu$-opioid receptor variants with sensitivity to pain. \\
\clearpage
\section*{Introduction}
Complex diseases are influenced by multiple environmental and genetic risk factors. A specific factor, such as a single mutation, may convey a high risk, but population frequencies of high risk factors are usually low, and substantial contribution to disease incidence can be attributable to accumulation of multiple but weak determinants within individuals. Genetic determinants of complex diseases that had been identified by genetic association studies tend to carry modest effects, yet power to detect such variants, as well as accuracy of identifying individuals at risk, can be improved by combining multiple weak predictors. The main challenge in detecting specific variants is low statistical power, but the overall accumulated effect of many individually weak signals can be much stronger. It is  convenient to combine statistical summaries of associations, for example, \Pvs{}, and this approach can be nearly as efficient as analysis of raw data.\cite{DanyuLinMetaNoGain2009} In observational research, methods for combining \Pvs{} are commonly associated with meta-analyses that pool results of multiple experiments studying the same hypothesis. The combined P-value then aggregates signals across all $L$ studies, potentially providing a higher level of assurance that the studied risk factor is associated with disease. Furthermore, if samples in those studies are taken from populations that are similar with respect to the effect size magnitude, the combined meta-analytic \Pv{} will well approximate the one that would have been obtained by pooling together all raw data and performing a single test.\cite{zaykin2011optimally}

\Pvs{} can also be combined when the $L$ hypotheses are distinct, and when the interest is in detecting the overall signal. Such applications are common and include gene set and pathway analyses. Specifically, a typical strategy in computation of gene- and pathway-scores includes (1) mapping individual SNPs to genes, followed by combining their association P-values into gene-scores, and (2) grouping genes into pathways and combining gene-scores into pathway-scores.

Existing tools for combining \Pv{}s ($P_i, i=1,\dots, L$) are often based on the sum of $P_i$'s transformed by some function $H$. For example, Fisher test\cite{fisher1932statistical} is based on the log-transformed \Pvs{}, $H(P_i) = -2 \ln(P_i)$, which are then added up to form a test statistics $T = \sum_{i=1}^{L} H(P_i) \sim \chi^2_{(L)}$, where $\chi^2_{(L)}$ has a chi-square distribution with $L$ degrees of freedom. When a portion of $L$ distinct associations is expected to be spurious, it is advantageous to combine only some of the predictors using a truncated variation of combined \Pv{} methods. For instance, Zaykin et al.\cite{Zaykin2002} proposed the Truncated Product Method (TPM) as a variation of the Fisher test, trimmed by the indicator function, $I(P_i \le \alpha)$, that is equal to zero if $P_i > \alpha$, and one if $P_i \le \alpha$; $0 < \alpha \le 1$ is a truncation threshold. The combined \Pv{}, $P_{\text{TPM}}$, is then given by the cumulative distribution function (CDF) of $W =  \sum_{i=1}^{L} \ln(P_i) I(P_i \le \alpha)$.
With the TPM approach, the threshold $\alpha$ is fixed while the number of \Pvs{} that form the sum $W$ is random. A related popular method for combining top-ranking \Pvs{} is the Rank Truncated Product (RTP).\cite{dudbridge2003,zaykin2007combining,ZaykinThesis} In RTP, the number of \Pvs{} to be combined, $k$, is fixed, rather than the \Pv{} threshold, as in TPM. The resulting combined \Pv{} can be found from the cumulative distribution of the product:
\begin{eqnarray*}
    P_{\text{RTP}} &=& \Pr \left\{ \prod_{i=1}^{k}  P_{(i)} \le w  \right\} 
  = 1 - \Pr \left\{ \sum_{i=1}^{k} \ln \left[P_{(i)}\right] > \ln \left[w\right] \right\},
\end{eqnarray*}
where $P_{(i)}$ is the $i$th smallest \Pv{}, $i = 1, \ldots, k$. RTP leads to an appealing extension, where $k$ can be chosen adaptively, to maximize statistical power.\cite{Hoh2001,Yu2009,zhang2013} Adaptive rank truncated product (aRTP) variations optimize selection of the truncation point $k$ among all (or a subset) of possible values $1 \le k \le L$. Adaptive extensions for TPM are not as straightforward because the threshold $\alpha$ is a continuous variable, but one can resort to evaluating the distribution over a set of grid points.\cite{sheng2013adaptive}  In adaptive extensions of TPM and RTP, the final test statistic is the minimum \Pv{} observed at various candidate truncation points.

The RTP null distribution is considerably more complicated than that of TPM.  Complexity of the RTP distribution is due to dependency between ordered \Pvs{}. When $k = L$, this dependency is inconsequential because a statistic is formed as a sum of $L$ terms and its value does not change if the terms are re-ordered. In fact, when $k=L$, the RTP \Pv{} is the same as the Fisher combined \Pv{}, derived via a CDF of a sum of independent chi-square variables. However, if $1 < k < L$, the $k$ smallest $P$-values remain correlated and dependent even if these $k$ values are randomly shuffled. The dependency is induced through $P_{(k+1)}$ being a random variable: when $P_{(k+1)}$ happens to be relatively small, the $k$ \Pvs{} have to squeeze into a relatively small interval from zero to that value. This induces positive dependency between random sets of $k$ smallest \Pvs{}, similar to the clustering effect in random effects models. Although the linear correlation can be eliminated by scaling the largest P-value, $P_{(k)}$, the $k$ values remain dependent, as illustrated in Figure \ref{fig:hole} (see ``Appendix (A-1)'' for discussion).

Applications of combining independent \Pvs{} remain important in statistical research, and there is clear preference among practitioners for methods that are based on simple and transparent approaches, such as the Fisher or the inverse normal (Stouffer's) tests.\cite{stouffer1949american,fisher1932statistical,zaykin2011optimally,whitlock2005combining,loughin2004systematic,won09} Here, we derive a simple, easily implemented theoretical from of the RTP distribution for independent \Pvs{} which further leads to derivation of a new statistic. The new statistic, which we call the Augmented RTP, or ART, is based on the product of the first smallest \Pvs{}, just like the RTP but, unlike the RTP, the distribution of the new statistic is given by standard functions and its computation avoids explicit integration.
Despite simplicity, ART is at least as powerful as RTP, according to our simulation studies. Moreover, the ART leads to an adaptive statistic, where the number of the smallest \Pvs{} to combine can be determined analytically to maximize power. Next, we extend our methods by allowing dependence in the observed \Pvs{}. In genetic association studies, \Pvs{} are often correlated due to linkage disequilibrium (LD). The LD correlation is typically accounted for through permutational or other resampling approaches, where \Pvs{} are simulated under the null hypothesis while preserving LD between genetic variants. While such approaches are practical and easy to implement, it is also possible to de-correlate \Pvs{} before combining them and then use any of the approaches developed under the independence assumption. Surprisingly, we find that the decorrelation step often improves power. In particular, we find that when association with disease is markedly different among variants within a gene, statistical power of standard methods (without the  decorrelation step) approaches a plateau as a function of LD and does not improve as the number of SNPs increases. In contrast, power of our proposed decorrelation method increases steadily with the number of SNPs. Our analytical results as well as simulation experiments demonstrate this property for both ART (where $k$ is chosen beforehand and fixed) and for the adaptive variations of RTP and ART (aRTP and ART-A). Finally, we illustrate usefulness of the proposed methods by strengthening an overall, gene-based association via combining previously reported \Pvs{} between pain sensitivity and individual SNPs within the $\mu$-opioid receptor.
\section*{Material and Methods}
\subsection*{Theoretical RTP distribution and Augmented RTP, the ART}
Even when \Pvs{} are independent, previously proposed theoretical forms of the RTP distribution are cumbersome and result in expressions that involve repeated integration.\cite{dudbridge2003,ZaykinThesis,CombPval07,Nagaraja2006} For example, Nagaraja\cite{Nagaraja2006} gives the cumulative distribution for the statistic $W_{k} = - \sum_{i=1}^k \ln P_{(i)}$ and $k<L$, as:
\begin{eqnarray}
   \Pr(W_k > w) &=& \sum_{j=1}^k w_j \exp\left\{-\frac{c_j w}{c_{k+1}}\right\} 
         \frac{1}{(L-k-1)!} \int_0^w \exp\left\{y \, d_j\right\} y^{L-k-1} d y \nonumber \\
          &+& \sum_{s=0}^{L-k-1} \exp\left\{ -w \right\} \frac{w^s}{s!},  \quad \text{where}\label{nagaraja} \\
   c_j &=& L-j+1, \nonumber\\
   d_j &=& \frac{k+1-j}{L-k}, \nonumber\\
   w_j &=& \frac{1}{L-j+1} \frac{L!}{(L-k)!} \frac{(-1)^k-j}{(j-1)!(k-j)!}. \nonumber
\end{eqnarray}
Theoretical forms of the RTP distribution (e.g., Eq. \ref{nagaraja}) may retain order-specific terms. Here, we proceed to a simpler representation by noting that every random realization of $k$ smallest \Pvs{} can be shuffled. This step does not change the value of  the product, $W_k$ (or its logarithm), which is our statistic of interest, but implies that we can treat the joint $k$-variate distribution as governed by the same pair-wise dependence for every pair of variables. Moreover, variables of that shuffled distribution are identical marginally. The dependency is induced completely through randomness of $P_{(k+1)}$, and conditionally on its value, the $\{W_k \mid p_{(k+1)}\}$ distribution is given by standard independence results. Then, $P_\text{RTP}$ is given by the marginal CDF of $W_k$. Based on this conceptual model, we derived the following representation of RTP where a single integral is evaluated in a bounded interval $(0,1)$, which allows one to evaluate the RTP distribution as a simple average of standard functions. Specifically, we derive a simple expression for the RTP distribution as the expectation of a function of a uniform (0 to 1) random variable:
\begin{eqnarray}
  P_{\text{RTP}}(k) &=& \Pr(W_{k} \le w) = 1-\int_0^1 G_{k}\left\{\ln\left(\frac{\left[B^{-1}_{k+1}(u)\right]^{k}}{w}\right)\right\} du \label{eq:wk}\\
  &=& E\left\{H(U \mid k, w)\right\}, \nonumber
\end{eqnarray}
where $B^{-1}_{k+1}(\cdot)$ is inverse CDF of $\text{Beta}(k+1, L-k)$ distribution, $G_k(\cdot)$ is CDF of Gamma$(k,1)$, and $H(u \mid k, w)=G_k\left(\ln \left(\frac{\left[B^{-1}_{k+1}(u)\right]^{k}}{w}\right)\right)$. $P_{\text{RTP}}(k)$ is the combined RTP \Pv{}. Notably, given the value of the product of $k$ \Pvs{}, $W=w$, we can simultaneously evaluate  $P_{\text{RTP}}(k+1)$:
\begin{eqnarray}
 P_{\text{RTP}}(k+1) &=& \Pr(W_{k+1} \le w) = 1-\int_0^1 G_{k}\left\{\ln\left(\frac{\left[B^{-1}_{k+1}(u)\right]^{k+1}}{w}\right)\right\} du
\label{eq:wk1}.
\end{eqnarray}
Details and the derivation are given in ``Appendix (A-2).''

The conditional independence of $k-1$ smallest \Pvs{}, given a value of the beta-distributed $k$-th smallest \Pv{} (Eq. \ref{eq:X}, \ref{eq:Y}), leads to a simple statistic which (just as RTP) is a function of of the product of the $k$-th smallest \Pvs{}. This statistic and its distribution are not an approximation to $W_k$ and the RTP distribution. However, similarly to RTP, the new statistic is designed to capture information contained in the first $k$ smallest $P$-values. To construct the new statistic, we propose the following transformation that involves the product $W_{k-1}$ and the variable $P_{(k)}$. These transformations yield three independent variables, that are next added together and give a gamma-distributed random variable,
\begin{eqnarray}
   A_k &=& -\ln \left\{ W_{k-1} \right\} + (k-1) \ln \left\{ P_{(k)} \right\} 
  + G_\lambda^{-1} \left\{1 - B_k(P_{(k)})\right\}, \label{ak.stat}
\end{eqnarray}
where $G_k^{-1}(\cdot)$ is inverse CDF of Gamma$(k,1)$, $$\lambda = (k-1) \times E\left\{-\ln\left(P_{(k)}\right)\right\} = (k-1) ( \Gamma' (L+1)/\Gamma (L+1) - \Gamma'(k)/ \Gamma(k)),$$ $\Gamma'$ is the first derivative of a gamma function; and $B_k(x)$ is the CDF of $\text{Beta}(k,L-k+1)$ distribution evaluated at $x$. The shape parameter $\lambda$ is chosen so that the two last terms in Eq. \ref{ak.stat} (that are both transformations of $P_{(k)}$) have the same expectation. Given the observed value $A_k = a_k$, the combined \Pv{} is
\begin{eqnarray}
   \text{ART} = \Pr(A_k \le a_k) = 1 - G_{k + \lambda - 1} (a_k). \label{ak}
\end{eqnarray}
Under the null hypothesis, as illustrated by Figure \ref{fig:Ak_RTP}, combined P-values based on the proposed method (ART) are very similar to $P_{\text{RTP}}$, and approach $P_{\text{RTP}}$ as $k$ increases. However, under the alternative, we find (as described in ``Results'' section) that ART has either the same or higher power than RTP. Furthermore, the combined \Pv{}, $\text{ART}$, can be easily computed in R using its standard functions. A short example and an R code are given in ``Appendix (A-3).''

\subsection*{Adaptive ART method, ART-A}
As we discussed earlier in Introduction, the number of $k$ \Pvs{} to be combined by the RTP method (or ART) is fixed and needs to be pre-specified. The choice of $k$ is somewhat arbitrary, so a researcher may wish to evaluate $\text{ART}$ at several values of $k$, consequently choosing $k$ that corresponds to the smallest combined \Pv{}. However, this additional step creates another layer of multiple comparisons, which needs to be accounted for. Yu et al. \cite{Yu2009} proposed an empirical procedure to evaluate adaptive RTP (aRTP) method based on the minimum \Pv{} computed over various candidate truncation points. To avoid a cumbersome two-level permutation procedure, they built on the method suggested by Ge et al.\citep{Ge2003} to reduce computational time. While computationally efficient, the method requires to store a large $B \times L$ matrix, with every row containing $L$ \Pvs{} generated under the null distribution over $B$ simulated experiments. Zhang et al.\cite{zhang2013} derived analytic but mathematically complex aRTP distribution, which needs to be evaluated using high-dimensional integration. Here, we propose a new and easily implemented version of the theoretical distribution for ART, ART-A. The method exploits the fact that ordered \Pvs{} can be represented as functions of the same number of independent uniform random variables (Appendix (A-4)). The two main ideas behind ART-A are: first to approximate the Gamma distribution with a large shape parameter by the normal distribution, and second to use the fact that the joint distribution of the partial normal sums follows a multivariate normal distribution.

\subsection*{Correlated $P$-values}
We further extend the proposed methods to combine correlated \Pvs{} via Decorrelation by Orthogonal Transformation approach, DOT. Let $L$ correlated $P$-values, $(p_1, p_2, \ldots, p_L)$, originate from statistics that jointly follow a multivariate normal disitrbution, $\mathbf{y} \sim \text{MVN}\left(\boldsymbol{\mu} = \matr{0}, \matr{\Sigma}\right)$, under $H_0$. Dependent variables can be transformed into independent variables by using eigendecomposition of $\matr{\Sigma}$, such that $\matr{\Sigma} = \matr{Q}\matr{\Lambda}\matr{Q^{-1}}$, where $\matr{Q}$ is a square matrix, with $i$th column containing eigenvector $\mathbf{q}_i$ of $\matr{\Sigma}$, and $\matr{\Lambda}$ is the diagonal matrix of eigenvalues $\lambda_1, \lambda_2, \ldots, \lambda_L$. Next, define an orthogonal matrix $\matr{H} = \matr{Q} \matr{\Lambda} ^{-1/2}\matr{Q}^T$ and $\matr{y}_e = \matr{H}^{T} \matr{y}$. \Pvs{} are decorrelated as $1 - \Phi^{-1} (\matr{y}_e)$. Then, the first $k$ smallest decorrelated \Pvs{} can be used to calculate various combined statistics. The choice of this particular transformation is motivated by its ``invariance to order'' property. Briefly, in the equicorrelation case, including the special case of $\rho=0$, i.e., independence, a permutation of $\matr{y}$ should yield the same (possibly permuted)  values in the decorrelated vector, $\matr{y}_e$. Extensive evaluation of the decorrelation approach are presented by is elsewhere. \cite{VegasPreprint}

\section*{Results}
\subsection*{Simulation study results}
We used simulation experiments to evaluate the Type I error rate and power of the proposed methods relative to the previously studied RTP (defined for a fixed $k$) and to the adaptive RTP (where $k$ is varied and the distribution is evaluated by single-layer simulations as in Yu et al, 2009).\cite{Ge2003,Yu2009} Performance of various methods was evaluated using $k$ first-ordered \Pvs{}, with $k=\{10, 100\}$ and $L = \{100, 200, 500\}$. Details of the simulation design are given in ``Appendix (A-5).''

Table \ref{tab1}-\ref{tabcor1} present Type I error rates for combinations of independent and decorrelated \Pvs{} respectively. In the tables, rows labeled ``ART-A'' refer to our newly proposed adaptive ART method, while ``aRTP'' rows label the results of the conventional approach.\cite{Yu2009}
For the adaptive methods, the sequence of truncation points varied from 1 to $k$ or from 1 to $L$, if $k=L$. Both tables confirm that all methods maintain the correct Type I error rate.

\begin{figure}[h!]
\centering
\includegraphics[width=0.3\linewidth]{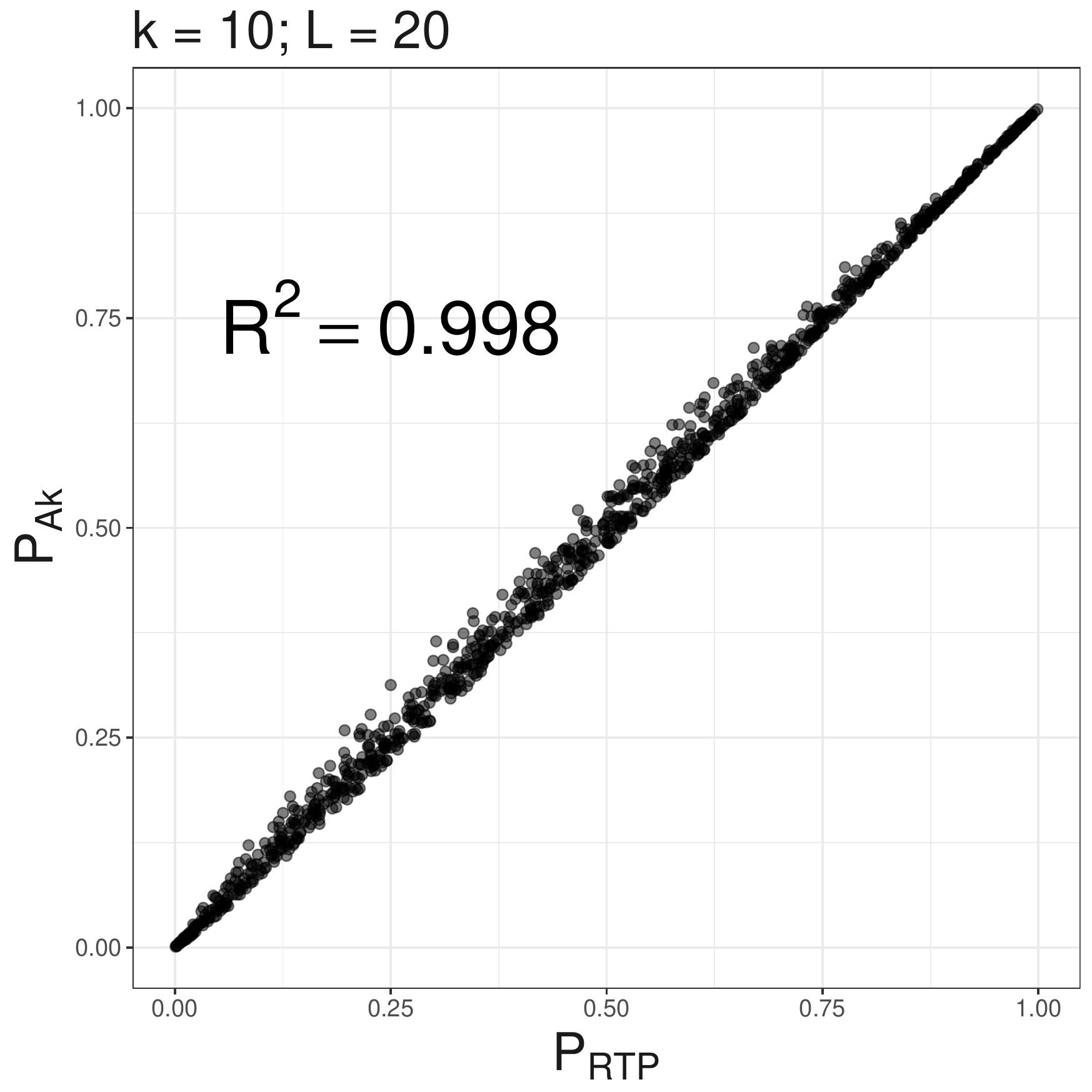} \quad \includegraphics[width=0.3\linewidth]{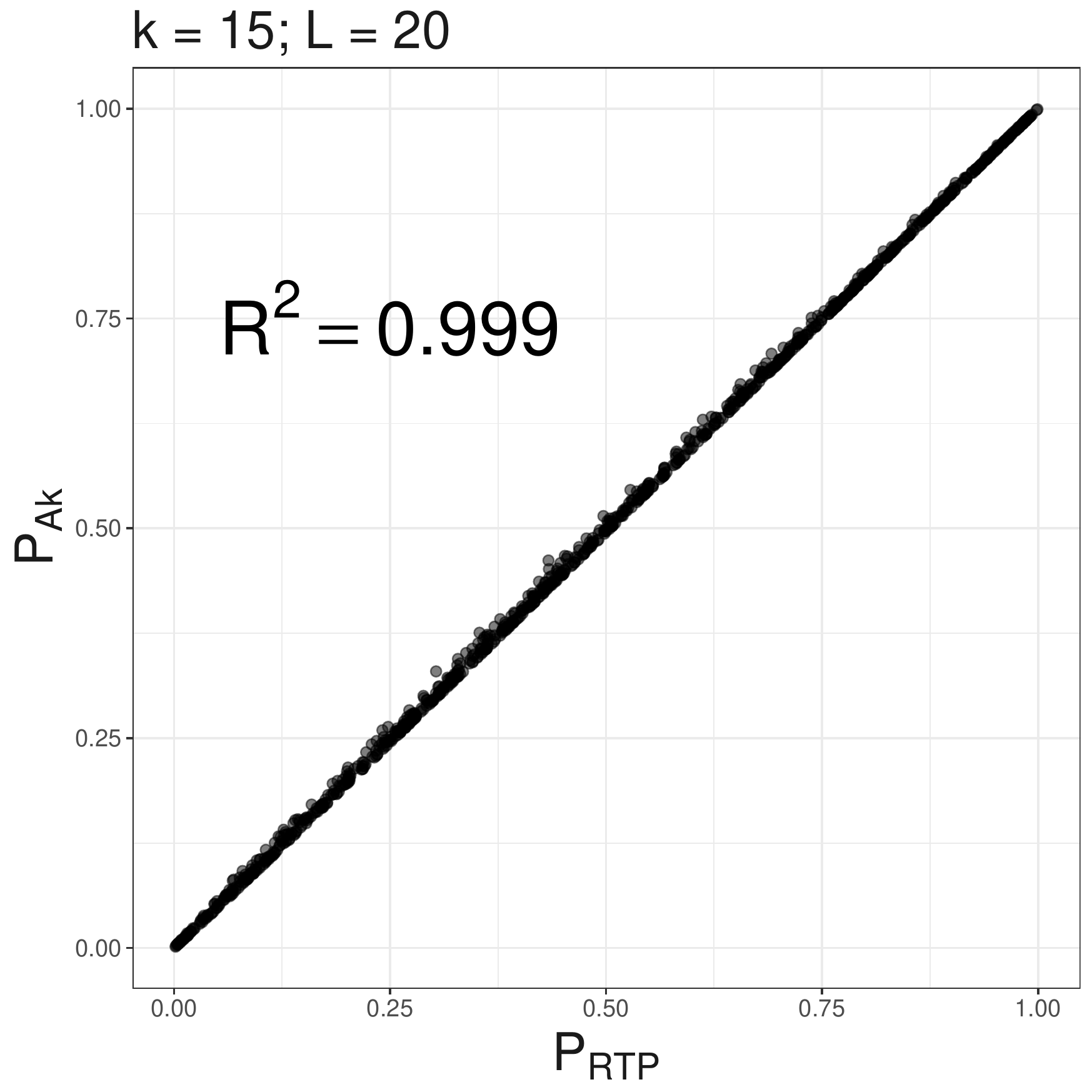} 
\caption{\textbf{Combined P-values based on A$_k$ versus RTP statistic.}\\ Multiple combined P-values were computed using the two proposed statistics based on either top 10 or top 15 P-values out of $L=20$ tests.
}
\label{fig:Ak_RTP}
\end{figure}

Tables \ref{tab2}-\ref{tab5} summarize a set of power simulations for independent \Pvs{}. Results presented in Table \ref{tab2} were obtained under the assumption that all $L$ statistics had the same underlying effect size ($\mu = 0.5$). From this table, it is evident that our ART has the highest power, closely followed by RTP. In general, the ART \Pvs{} tend to be similar to the \Pvs{} obtained by the RTP, and we show their similarity graphically in Figure \ref{fig:Ak_RTP}. The Simes method has the lowest power, which is expected due to homogeneity in effect sizes across $L$ tests and absence of true nulls. For the results in Table \ref{tab3}, the effect size was allowed to randomly vary throughout the range from 0.05 to 0.45. In both of these tables, the ART method has the highest power, while the Simes method has the lowest power. The power of both adaptive methods is very similar to one another but lower than that of methods based on a fixed $k$ (RTP and ART). Nonetheless, in practice, a good choice for $k$ may not be immediately clear, so a small sacrifice in power may be preferable to an arbitrary and possibly poor choice of $k$. However, when $L$ is large, it can be impractical or unreasonable to vary candidate truncation points all the way up to $L$. Finally, Table \ref{tab5} summarizes results for simulations when some of the $L$ hypotheses were true nulls ($\mu=0$), while the remaining hypotheses were true signals ($\mu=0.5$). The results follow the same pattern as in the previous tables, with ART having the highest power.

Table \ref{tabcor4} summarizes a set of power simulations for correlated \Pvs{}. The effect sizes were randomly varied between -0.45 and 1.3 in each simulation. The correlation matrices were generated as described in ``Appendix (A-5).'' This set of simulations assumes that the \Pvs{} were obtained from the same data set as the sample estimate of the correlation matrix. Under heterogeneous effect sizes (Table \ref{tabcor4}) the empirical versions of the tests (``RTP'', ``ART-A'') show nearly identical (and low) power for various combinations of $k$ and $L$ values. However, decorrelation-based methods become quite powerful, and their power is increasing with $k$ and $L$. The steady power increase is due to the decorrelation effect on the combined noncentrality that involves the sum  $\sum^L_{i \ne j} (\mu_i - \mu_j)^2$, which increases with the increased heterogeneity of $\mu$. More details on the performance of the decorrelation approach are given by us elsewhere,\cite{VegasPreprint} but here we briefly note that this finding is practically relevant because substantial heterogeneity of associations is expected among genetic variants, leading to a substantial power boost, as we next illustrate via re-analysis of published associations of genetic variants within the $\mu$-opioid gene with
pain sensitivity.

\subsection*{Real data analysis}
In several popular variations of the gene-based approach,\cite{neale2004future} association statistics or \Pvs{} are combined across variants within a gene.\cite{Yu2009,peng2010gene,liu2010versatile,li2011gates} Gene-based approaches have some advantages over methods based on individual SNPs or haplotypes. In particular, gene-based \Pvs{} may facilitate subsequent meta-analysis of separate studies and can be less susceptible to erroneous findings.\cite{neale2004future} To obtain a gene-based \Pv{}, one would need to account for LD among variants. The matrix of LD correlation coefficients can be obtained conveniently without access to individual genotypes if frequencies of haplotypes for SNPs within the genetic region of interest are available. The LD for alleles $i$ and $j$ is defined by the difference between the di-locus haplotype frequency, $P_{ij}$, and the product of the frequencies of two alleles, $D_{ij} = P_{ij} - p_{i}p_{j}$. The LD-correlation for SNPs $i$ and $j$ is $r_{ij} = \frac{D_{ij}}{p_i(1-p_i)p_j(1-p_j)}$. Shabalina et al.\cite{shabalina2008expansion} and Kuo et al. \cite{kuo2014discovering} reported SNP-based \Pvs{} (Table \ref{SNP.Pval}), as well as results of several haplotype-based tests for genetic association of variants within the $\mu$-opioid receptor (\textit{MOR}) with pain sensitivity. Kuo et al. also reported estimated frequencies for 11-SNP haplotypes within the \textit{MOR} gene,\cite{kuo2014discovering} from which the $11 \times 11$ LD correlation matrix can be computed. The $P_{ij}$ frequencies are given by the sum of frequencies of those 11-SNP haplotypes  that contain both of the minor alleles for SNPs $i$ and $j$. Similarly, $p_i$ allele frequency is the sum of haplotype frequencies that carry the minor allele of the SNP $i$. The LD correlations within the \textit{MOR} region spanned by the 11 SNPs ranged from -0.82 to 0.99, with the average absolute value $\approx 0.55$ and the median absolute value $\approx 0.66$. Half of pairwise LD correlations were smaller than -0.23 or larger than 0.82. Our analysis (Table \ref{muopioid}) showcases utility of the proposed methods. The columns show combined P-values, for $k$ varying from two up to all eleven SNPs ($k$=1 is equivalent to the Bonferroni correction, i.e., 0.007$\times$11). Similar to what we found via simulation experiments, where correlation is controlled by reshuffling the phenotype values while keeping the original LD structure intact, RTP and aRTP (without the decorrelation step) do not benefit from inclusion of additional SNPs. \Pvs{} in the ART column  are very similar to those in the RTP column, which reassures our theoretical expectations. In contrast to previously proposed methods that control correlation by resampling (i.e.,  RTP, aRTP and ART), the results in columns marked by ``decorr'' are substantially lower. In these columns, we used the proposed transformation to independence, which gives much stronger combined \Pvs{}. In all ``decorr'' columns, $k$=7 results in the smallest combined \Pv{}, implying that the number of real effects (including proxy associations) is at least seven.
\section*{Discussion}
Complex diseases are influenced by collective effects of environmental exposures and genetic determinants.  There can be numerous weak but biologically meaningful risk factors. The challenge is to distinguish between real and spurious statistical signals in the presence of multiple comparisons and low detection power. When the number of potential real associations is expected to be small, compared to the total number of variants evaluated within a study, it is advantageous to focus on the top-ranking associations. The rank truncated product method (RTP) has been designed with this objective in mind. The RTP and related approaches had been shown to be valuable tools in analysis of genetic associations with disease. In this article, we derive a mathematically simple form of the RTP distribution that leads a to new method, ART and its adaptive version, ART-A, that searches through a number of candidate values of truncation points and finds an optimal number in terms of combined \Pv{}. The ART is designed with the same objectives in mind as RTP and TPM: to facilitate detection of possibly weak signals among top-ranking predictors that could have been missed, unless combined into a single score. The ART is trivial to implement in terms of standard functions, provided by packages such as R, and its power characteristics are close to RTP or higher in all studied settings. Analytical forms of ART and ART-A are derived under independence. To accommodate LD, we propose a decorrelation step, by transformation of \Pvs{} to independence. Our Decorrelation by Orthogonal Transformation approach (DOT) is analogous to the Mahalanobis transformation.\cite{hardle2007applied} We found DOT to be surprisingly powerful in many settings, compared to the usual method of evaluating the distribution of product of correlated \Pvs{} under the null hypothesis. Theoretical properties and extensive numerical evaluation of DOT will be published elsewhere and currently these findings are available as a preprint.\cite{VegasPreprint} Further, we illustrate an application of our methods with analyses of variants within the $\mu$-opioid gene that have been shown to affect sensitivity to pain. We find strengthened evidence of overall association within the 11-SNP block. In this application, the LD correlation matrix was reconstructed from the haplotype frequencies, which  might be slightly different from the correlation of (0,1,2) values between pairs of SNPs.\cite{zaykin2004bounds} Further studies are needed to investigate whether approaches such as this, or utilization of reference panel (external) data as a source of LD information, may lead to substantial bias.

\section*{Declaration of Interests}
The authors declare no competing interests.

\section*{Acknowledgments}
This research was supported in part by the Intramural Research Program of the NIH, National Institute of Environmental Health Sciences.

\section*{Web Resources}
The URL for software referenced in this article is available at: \mbox{}\\
\noindent
{\small\url{https://github.com/dmitri-zaykin/Total_Decor}}

\clearpage
\bibliography{ART_05_02_19}

\begin{thebibliography}{10}

\bibitem{DanyuLinMetaNoGain2009}
Lin D, Zeng D.
\newblock {Meta-analysis of genome-wide association studies: no efficiency gain
  in using individual participant data}.
\newblock {Genet Epidemiol}. 2010;34(1):60--66.

\bibitem{zaykin2011optimally}
Zaykin DV.
\newblock {Optimally weighted Z-test is a powerful method for combining
  probabilities in meta-analysis}.
\newblock J Evol Biol. 2011;24(8):1836--1841.

\bibitem{fisher1932statistical}
Fisher SRA.
\newblock {Statistical methods for research workers}.
\newblock Genesis Publishing Pvt Ltd; 1932.

\bibitem{Zaykin2002}
Zaykin DV, Zhivotovsky LA, Westfall PH, Weir BS.
\newblock Truncated product method for combining P-values.
\newblock {Genet Epidemiol}. 2002;22(2):170--85.

\bibitem{dudbridge2003}
Dudbridge F, Koeleman BP.
\newblock Rank truncated product of P-values, with application to genomewide
  association scans.
\newblock {Genet Epidemiol}. 2003;25(4):360--366.

\bibitem{zaykin2007combining}
Zaykin DV, Zhivotovsky LA, Czika W, Shao S, Wolfinger RD.
\newblock Combining p-values in large-scale genomics experiments.
\newblock Pharmaceutical statistics. 2007;6(3):217--226.

\bibitem{ZaykinThesis}
Zaykin DV.
\newblock {Statistical analysis of genetic associations}.
\newblock Ph.D. thesis. North Carolina State University; 2000.

\bibitem{Hoh2001}
Hoh J, Wille A, Ott J.
\newblock Trimming, weighting, and grouping SNPs in human case-control
  association studies.
\newblock Genome Res. 2001;11(12):2115--9.

\bibitem{Yu2009}
Yu K, Li QZ, Bergen AW, Pfeiffer RM, Rosenberg PS, Caporaso N, et~al.
\newblock Pathway analysis by adaptive combination of P-Values.
\newblock {Genet Epidemiol}. 2009;33(8):700--709.

\bibitem{zhang2013}
Zhang S, Chen HS, Pfeiffer RM.
\newblock A combined p-value test for multiple hypothesis testing.
\newblock Journal of Statistical Planning and Inference. 2013;143(4):764--770.

\bibitem{sheng2013adaptive}
Sheng X, Yang J.
\newblock An adaptive truncated product method for combining dependent
  p-values.
\newblock Economics letters. 2013;119(2):180--182.

\bibitem{stouffer1949american}
Stouffer SA, DeVinney LC, Suchmen EA.
\newblock {The American soldier: Adjustment during army life}. vol.~1.
\newblock Princeton University Press, Princeton, US; 1949.

\bibitem{whitlock2005combining}
Whitlock MC.
\newblock {Combining probability from independent tests: the weighted Z-method
  is superior to Fisher's approach.}
\newblock Journal of Evolutionary Biology. 2005;18(5):1368--1373.

\bibitem{loughin2004systematic}
Loughin TM.
\newblock A systematic comparison of methods for combining p-values from
  independent tests.
\newblock Computational statistics \& data analysis. 2004;47(3):467--485.

\bibitem{won09}
Won S, Morris N, Lu Q, Elston RC.
\newblock {Choosing an optimal method to combine P-values}.
\newblock Statistics in medicine. 2009;28(11):1537--1553.

\bibitem{CombPval07}
Zaykin DV, Zhivotovsky LA, Czika W, Shao S, Wolfinger RD.
\newblock {Combining P-values in large-scale genomics experiments.}
\newblock Pharm Stat. 2007;6(3):217--226.

\bibitem{Nagaraja2006}
Nagaraja HN.
\newblock Order statistics from independent exponential random variables and
  the sum of the top order statistics.
\newblock In: Advances in Distribution Theory, Order Statistics, and Inference.
  Springer; 2006. p. 173--185.

\bibitem{Ge2003}
Ge Y, Dudoit S, Speed TP.
\newblock Resampling-based multiple testing for microarray data analysis.
\newblock Test. 2003;12(1):1--77.

\bibitem{VegasPreprint}
Vsevolozhskaya OA, Shi M, Hu F, Zaykin DV. Gene-set analysis by combining
  decorrelated association statistics; 2019.
\newblock \url{https://arxiv.org/abs/1906.02321}.

\bibitem{neale2004future}
Neale BM, Sham PC.
\newblock The future of association studies: gene-based analysis and
  replication.
\newblock The American Journal of Human Genetics. 2004;75(3):353--362.

\bibitem{peng2010gene}
Peng G, Luo L, Siu H, Zhu Y, Hu P, Hong S, et~al.
\newblock Gene and pathway-based second-wave analysis of genome-wide
  association studies.
\newblock European Journal of Human Genetics. 2010;18(1):111.

\bibitem{liu2010versatile}
Liu JZ, Mcrae AF, Nyholt DR, Medland SE, Wray NR, Brown KM, et~al.
\newblock A versatile gene-based test for genome-wide association studies.
\newblock The American Journal of Human Genetics. 2010;87(1):139--145.

\bibitem{li2011gates}
Li MX, Gui HS, Kwan JS, Sham PC.
\newblock GATES: A Rapid and Powerful Gene-Based Association Test Using
  Extended Simes Procedure.
\newblock The American Journal of Human Genetics. 2011;88(3):283--293.

\bibitem{shabalina2008expansion}
Shabalina SA, Zaykin DV, Gris P, Ogurtsov AY, Gauthier J, Shibata K, et~al.
\newblock Expansion of the human $\mu$-opioid receptor gene architecture: novel
  functional variants.
\newblock Human molecular genetics. 2008;18(6):1037--1051.

\bibitem{kuo2014discovering}
Kuo CL, Diatchenko L, Zaykin D.
\newblock Discovering multilocus associations with complex pain phenotypes.
\newblock Pain Genetics: Basic to Translational Science. 2014;p. 99--114.

\bibitem{hardle2007applied}
H{\"a}rdle W, Simar L.
\newblock Applied multivariate statistical analysis. vol. 22007.
\newblock Springer; 2007.

\bibitem{zaykin2004bounds}
Zaykin DV.
\newblock Bounds and normalization of the composite linkage disequilibrium
  coefficient.
\newblock {Genet Epidemiol}. 2004;27(3):252--257.

\bibitem{balakrishnan1998order}
Balakrishnan N, Rao CR.
\newblock Order statistics: theory \& methods.
\newblock Elsevier Amsterdam; 1998.

\bibitem{mi2009mvtnorm}
Mi X, Miwa T, Hothorn T.
\newblock mvtnorm: New numerical algorithm for multivariate normal
  probabilities.
\newblock The R Journal. 2009;1(1):37--39.

\bibitem{bartlett1951}
Bartlett MS.
\newblock An inverse matrix adjustment arising in discriminant analysis.
\newblock The Annals of Mathematical Statistics. 1951;22(1):107--111.

\bibitem{simes1986}
Simes R.
\newblock {An improved Bonferroni procedure for multiple tests of
  significance}.
\newblock Biometrika. 1986;73(3):751--754.

\bibitem{sidak1967}
\v{S}id\'{a}k Z.
\newblock Rectangular confidence regions for the means of multivariate normal
  distributions.
\newblock Journal of the American Statistical Association. 1967;78:626--633.

\bibitem{bonferroni1935calcolo}
Bonferroni CE.
\newblock Il calcolo delle assicurazioni su gruppi di teste.
\newblock Studi in onore del professore salvatore ortu carboni. 1935;p. 13--60.

\bibitem{benjamini1995cfd}
Benjamini Y, Hochberg Y.
\newblock {Controlling the false discovery rate: a practical and powerful
  approach to multiple testing}.
\newblock {J Royal Stat Soc B}. 1995;57(1):289--300.

\end{thebibliography}
\clearpage
\section*{Figure Titles and Legends}
\begin{figure}[h!]
\centering
 \includegraphics[width=0.3\linewidth]{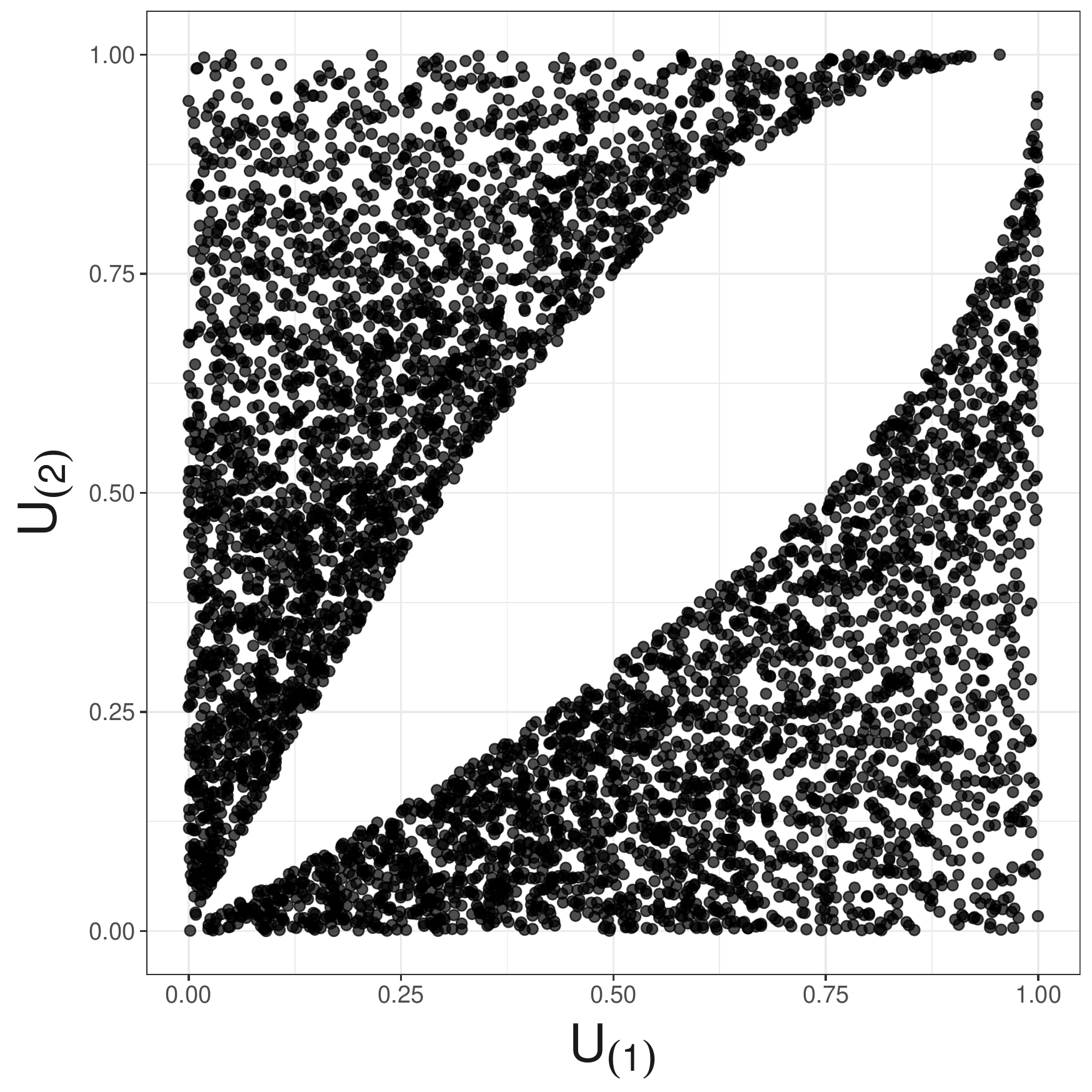} 
\caption{\textbf{Illustration for decorrelated yet dependent \Pvs{}; $\mathbf{k=2}$, $\mathbf{n=4}$.}\\ A plot of simulated and decorrelated values, $U_{(1)}$ vs $U_{(2)}$, reveals a hole in the middle, instead of the complete Malevich black square, indicating dependency.
}
\label{fig:hole}
\end{figure}

\clearpage
\section*{Tables}
\begin{table}[h]
  \centering
  \begin{tabular}{rrrrrrr}
    \hline
    & \multicolumn{3}{c}{$k=10$} & \multicolumn{3}{c}{$k= 100$ }\\ 
    \hline
    & $L$ = 100 & $L$ = 200 & $L$ = 500 & $L$ = 100 & $L$ = 200 & $L$ = 500 \\ 
    \hline
   RTP           & 0.05 & 0.05 & 0.05 & 0.05 & 0.05 & 0.05 \\ 
   ART & 0.05 & 0.05 & 0.05 & 0.05 & 0.05 & 0.05 \\ 
   aRTP          & 0.05 & 0.05 & 0.05 & 0.05 & 0.05 & 0.05 \\ 
   ART-A     & 0.05 & 0.05 & 0.05 & 0.05 & 0.05 & 0.05 \\ 
   Simes         & 0.05 & 0.05 & 0.05 & 0.05 & 0.05 & 0.05 \\ 
   \hline
  \end{tabular}
  \caption{Type I error at $\alpha = 0.05$, assuming that \Pvs{} are independent.} \label{tab1}
\end{table}
\begin{table}[!htbp]
	\centering
	\begin{tabular}{rrrrrr}
		\hline
		& $L$ = $k$ = 4 & $L$ = $k$ = 6 & $L$ = $k$  = 10 & $L$ = 100, $k$ = 10 & $L$ = $k$ = 100 \\ 
		\hline
		Mean $|\rho|$ & 0.39 & 0.43 & 0.45 & 0.47 & 0.47 \\
		\hline
		RTP                    & 0.05 & 0.05 & 0.05 & 0.05 & 0.05 \\ 
		RTP(decorr)            & 0.05 & 0.05 & 0.05 & 0.05 & 0.05 \\ 
		ART(decorr)  & 0.05 & 0.05 & 0.05 & 0.05 & 0.05 \\ 
		aRTP                   & 0.05 & 0.05 & 0.05 & 0.05 & 0.05 \\ 
		ART-A(decorr)      & 0.05 & 0.05 & 0.05 & 0.05 & 0.05 \\ 
		Simes                  & 0.05 & 0.05 & 0.05 & 0.05 & 0.05 \\ 
		\hline
	\end{tabular}
	\caption{Type I error at $\alpha = 0.05$ for randomly correlated \Pvs{}.} \label{tabcor1}
      \end{table}
\begin{table}[h]
  \centering
  \begin{tabular}{rrrrrrr}
    \hline
    & \multicolumn{3}{c}{$k=10$} & \multicolumn{3}{c}{$k= 100$ }\\ 
    \hline
    & $L$ = 100 & $L$ = 200 & $L$ = 500 & $L$ = 100 & $L$ = 200 & $L$ = 500 \\ 
    \hline
    RTP                   & 0.35 & 0.43 & 0.54 & 0.49 & 0.73 & 0.94 \\ 
    ART         & 0.38 & 0.49 & 0.63 & 0.50 & 0.74 & 0.95 \\ 
    aRTP                  & 0.27 & 0.33 & 0.41 & 0.41 & 0.61 & 0.86 \\ 
    ART-A             & 0.32 & 0.38 & 0.46 & 0.40 & 0.57 & 0.72 \\ 
    Simes                 & 0.14 & 0.16 & 0.17 & 0.15 & 0.16 & 0.17 \\ 
    \hline
  \end{tabular}
  \caption{Power under the alternative hypothesis, assuming independence and the same effect size $\mu = 0.5$ for all $L$ tests.} \label{tab2}
\end{table}
\begin{table}[h]
  \centering
  \begin{tabular}{rrrrrrr}
    \hline
    & \multicolumn{3}{c}{$k=10$} & \multicolumn{3}{c}{$k= 100$ }\\ 
    \hline
    & $L$ = 100 & $L$ = 200 & $L$ = 500 & $L$ = 100 & $L$ = 200 & $L$ = 500 \\ 
    \hline 
    RTP            & 0.23 & 0.28 & 0.37 & 0.30 & 0.46 & 0.72 \\
    ART  & 0.25 & 0.32 & 0.43 & 0.30 & 0.46 & 0.75 \\
    aRTP           & 0.18 & 0.22 & 0.27 & 0.25 & 0.36 & 0.57 \\
    ART-A      & 0.22 & 0.26 & 0.33 & 0.25 & 0.37 & 0.53 \\
    Simes          & 0.12 & 0.13 & 0.14 & 0.12 & 0.13 & 0.14 \\
    \hline
  \end{tabular}
  \caption{Power under the alternative hypothesis, assuming independence and random effect size ($\mu$ between 0.05 and 0.45).} \label{tab3}
\end{table} 
\begin{table} [!htbp]
  \centering
  \begin{tabular}{rrrrrrr}
    \hline
    & \multicolumn{3}{c}{$k=10$} & \multicolumn{3}{c}{$k=50$} \\ 
    \hline
   Proportion of true effects & 2.5\% & 5\% & 10\% & 2.5\% & 5\% & 10\% \\ 
    \hline
  RTP             & 0.24 & 0.48 & 0.83 & 0.29 & 0.65 & 0.97 \\ 
  ART    & 0.24 & 0.52 & 0.89 & 0.29 & 0.66 & 0.98 \\ 
  aRTP             & 0.20 & 0.40 & 0.75 & 0.25 & 0.55 & 0.93 \\ 
  ART-A        & 0.22 & 0.45 & 0.75 & 0.26 & 0.56 & 0.88 \\ 
  Simes            & 0.14 & 0.23 & 0.38 & 0.14 & 0.23 & 0.38 \\ 
  \hline
  \end{tabular}
  \caption{Power under independence, assuming constant effect size ($\mu = 1.4$) for a fraction of $L$=1000 hypotheses and $\mu=0$ for the rest of the tests.} \label{tab5} 
\end{table}
\begin{table}[!htbp]
	\centering
	\begin{tabular}{rrrrrr}
		\hline
		& $L$ = $k$ = 4 & $L$ = $k$ = 6 & $L$ = $k$  = 10 & $L$ = 100, $k$ = 10 & $L$ = $k$ = 100 \\ 
		\hline
        Mean $|\rho|$ & 0.39 & 0.43 & 0.45 & 0.47 & 0.47  \\ 
		\hline
RTP                     & 0.13 & 0.12 & 0.11 & 0.17 & 0.12 \\ 
RTP(decorr)             & 0.41 & 0.47 & 0.57 & 0.98 & $>$0.99 \\ 
ART(decorr)   & 0.41 & 0.47 & 0.57 & 0.99 & $>$0.99 \\ 
aRTP                    & 0.17 & 0.16 & 0.16 & 0.20 & 0.18 \\ 
ART-A(decorr)       & 0.38 & 0.44 & 0.52 & 0.94 & 0.98 \\ 
Simes                   & 0.35 & 0.38 & 0.41 & 0.63 & 0.64 \\ 
\hline
	\end{tabular}
        \caption{Power for correlated \Pvs{} when the effect size is randomly varied between -0.45 and 1.3.} \label{tabcor4}
\end{table}
\begin{table}[ht]
\centering
\begin{tabular}{rrrrrr}
  \hline
   \multicolumn{6}{c}{$L=11$, mean $|\rho|=0.55$} \\
  $k$: & 5 & 6 & 7 & 9 & 11 \\
    \hline
  RTP                     & 0.07 & 0.06 & 0.06 & 0.05 & 0.05 \\ 
  RTP(decorr)             & 0.85 & 0.87 & 0.87 & 0.87 & 0.87 \\ 
  ART(decorr)     & 0.85 & 0.87 & 0.87 & 0.87 & 0.87 \\ 
  aRTP                    & 0.07 & 0.07 & 0.06 & 0.06 & 0.06 \\ 
  ART-A(decorr)        & 0.85 & 0.85 & 0.85 & 0.84 & 0.83 \\ 
  Simes                   & 0.78 & 0.78 & 0.78 & 0.78 & 0.79 \\ 
   \hline
\end{tabular}
\caption{Power at $\alpha = 0.05$ for \Pvs{} correlated according to the LD structure in the $\mu$-opioid gene, with effect sizes randomly sampled in the interval from -0.5 to 0.2.} \label{TabPowerMuOpioid} 
\end{table}
\begin{table}[!htbp]
	\centering
	\begin{tabular}{lr}
		\hline
		SNP & P-value \\ 
		\hline
		rs563649 & 0.0007 \\ 
		rs9322446 & 0.0941 \\ 
		rs2075572 & 0.2957 \\ 
		rs533586 & 0.7037 \\ 
		rs540825 & 0.8171 \\ 
		rs675026 & 0.8012 \\ 
		rs660756 & 0.5745 \\ 
		rs677830 & 0.9891 \\ 
		rs623956 & 0.8308 \\ 
		rs609148 & 0.8208 \\ 
		rs497332 & 0.3139 \\ 
		\hline
	\end{tabular}
	\caption{Individual SNP \Pvs{} as originally reported in Shabalina et al. \cite{shabalina2008expansion}} \label{SNP.Pval}
\end{table}
\begin{table}[ht]
	\centering
	\begin{tabular}{rrrrrrr}
		\hline
		k & RTP & RTP (decorr) & ART (decorr) & aRTP & ART-A (decorr) \\ 
		\hline
		2 & 0.0187 & 0.0225   & 0.0256 & 0.0519  & 0.0533  \\ 
		3 & 0.0411 & 0.0234   & 0.0253 & 0.2963  & 0.0211  \\ 
		4 & 0.0566 & 0.0192   & 0.0183 & 0.1845  & 0.0115  \\ 
		5 & 0.0886 & 0.0211   & 0.0231 & 0.4702  & 0.0165  \\ 
		6 & 0.1172 & 0.0204   & 0.0208 & 0.6543  & 0.0070  \\ 
		7 & 0.1486 & \textbf{0.0177}   & \textbf{0.0169} & 0.7718  & \textbf{0.0041}  \\ 
		8 & 0.1726 & 0.0211   & 0.0220 & 0.7189  & 0.0416  \\ 
		9 & 0.1810 & 0.0228   & 0.0232 & 0.6766  & 0.0165  \\ 
		10 & 0.1867 & 0.0241  & 0.0241 & 0.6423  & 0.0096  \\ 
		11 & 0.1938 & 0.0241  & 0.0243 & 0.6140  & 0.0065  \\ 
		\hline
	\end{tabular}
   \caption{Combined \Pvs{} by different methods for $\mu$-opioid data. The smallest \Pvs{} in ``decor'' columns are highlighted in bold. The table reports 11 aRTP values rather than a single optimal one, because one can specify the largest $k$ value, which was varied here from 2 to 11.
}\label{muopioid} 
\end{table}

\clearpage
\appendix
\doublespacing
\section*{Appendix}
\renewcommand{\thesubsection}{A-\arabic{subsection}}
\setcounter{equation}{0}
\renewcommand{\theequation}{A-\arabic{equation}}

\subsection{Correlation and dependencies among $k$ smallest \Pvs{}}
The complexity of analytic forms of the RTP distribution is due to dependency introduced by ordering of $P$-values. Although order statistics are correlated, products and sums are oblivious to the order of the terms, therefore for the case when $k=L$, the statistic $T_k$ follows the gamma distribution with the shape parameter equal to $L$, and the unit scale, i.e., $T_L \sim \text{Gamma}(L,1)$. This is essentially the same as the Fisher combined \Pv{}, where the statistic is $2T_L$, distributed as the chi-square with $2L$ degrees of freedom. However, for $1 \le k < L$, the $k$ smallest $P$-values remain dependent even if these $k$ values are not sorted (e.g., randomly shuffled). The dependency is induced through $P_{(k+1)}$ being a random variable: when $P_{(k+1)}$ happens to be relatively small, the $k$ \Pvs{} have to squeeze into a relatively small interval from zero to that value. This induces a positive correlation between random sets of $k$ smallest \Pvs{}, similar to the clustering effect in the random effects models. 

The $k$ smallest unordered \Pvs{} are equicorrelated  and also have the same marginal distribution, which can be obtained as a permutation distribution of the first $k$ uniform order statistics. Assuming independence of $L$ \Pvs{} and their uniform distribution under the null hypothesis, we can derive the correlation between any pair of {\em unordered} $k$ smallest \Pvs{} as  $\rho(k,L) = 3(L-k) / (2 + k(L-2) + 5L)$.  As $L$ increases, the correlation approaches the limit that no longer depends on $L$: $\lim_{L\rightarrow \infty} \rho(k,L) = 3/(k+5)$. The correlation can be substantial for small $k$ and cannot be ignored. There is a very simple transformation that makes a set of $k$ \Pvs{} uncorrelated. All that is needed to decorrelate these \Pvs{} is to scale the largest of them:
\begin{eqnarray}
   X_1 &=& P_{(1)} \nonumber \\
   X_2 &=& P_{(2)} \nonumber \\
   & \vdots & \nonumber \\
   X_{k-1} &=& P_{(k-1)} \nonumber \\
  X_{k} &=& \sigma P_{(k)}, \nonumber
\end{eqnarray}
where
\begin{eqnarray}
   \sigma = \frac{2 L - k + 3 + \sqrt{(k+1)(L+1)(L-k+1)}}{4 + 2 L}, \nonumber
\end{eqnarray}
and then randomly shuffle the set $X_1,\dots X_k$. This scale factor $\sigma$ can be derived by solving the mixture covariance linear equations induced by the permutation distribution of the first $k$ order statistics. The decorrelated values can be further transformed so that each has the uniform (0,1) distribution marginally:
\begin{eqnarray}
 U_j &=& \frac{1}{k} \sum_{i=1}^k \text{Beta}(X_j; \,\,i, L-i+1), \,\,\,\,j=1,\dots,k-1 \\
 U_k &=&  \frac{1}{k} \sum_{i=1}^k \text{Beta}(X_k/\sigma; \,\,i, L-i+1),
\end{eqnarray}
where Beta$(x; \,\, a,b)$ is the CDF of a beta$(a,b)$ distribution evaluated at $x$. Although the scaling and subsequent shuffle removes the correlation, the values remain dependent, as illustrated in Figure \ref{fig:hole}.

\subsection{Derivation of the RTP distribution}
An intuitive way to understand our derivation of the RTP distribution is through references to simulations. The simplest, brute-force algorithm to obtain the RTP combined \Pv{} is by simulating its distribution directly. If $w_k$ is the product of $k$ actual \Pvs{}, one can repeatedly ($B$ times) simulate $L$ Uniform(0,1) random variables $U_i$, sort them, take the product of $k$ smallest values, and compare the resulting product to $w_k$. As the number of simulations, $B$, increases, the proportion of times that simulated values will be smaller than $w_k$ converges to the true combined RTP \Pv{}.

There are several ways to optimize the above simulation scenario with respect to computational complexity. For instance, sets of ordered uniform \Pvs{} can be simulated directly using well-known results from the theory of order statistics. Despite the fact that the marginal distribution of $i$th ordered value is Beta$(i, L-i+1)$, to create the necessary dependency between the ordered \Pvs{}, sets of $k$ values have to be simulated in a step-wise, conditional fashion. The minimum value, $P_{(1)}$, can be sampled from Beta$(1, L)$ distribution. Alternatively, using the relationship between beta and Uniform(0,1) random variables, it can be sampled as $P_{(1)} = 1-U_1^{1/L}$. Next, since the value $P_{(2)}$ cannot be smaller than $P_{(1)} = p_{(1)}$, conditionally on the obtained value, it has to be generated from a truncated beta distribution. The third smallest value should be sampled conditionally on the second one, and so on.\cite{balakrishnan1998order} Therefore, the sequence and the product $w_k$ can be obtained by simulating $k$ ordered \Pvs{}, rather then all $L$ unsorted values.
\begin{equation} 
\begin{split}
 P_{(1)}  &=  1 - U_1^{\frac{1}{L}}  \\
 P_{(2)}  &=  1 -  u_1^{\frac{1}{L}} U_2^{\frac{1}{L-1}} \\
 P_{(3)}  &=  1 -  u_1^{\frac{1}{L}} u_2^{\frac{1}{L-1}} U_3^{\frac{1}{L-2}}  \\
\vdots \\
 P_{(k)}  &=  1-u_1^{\frac{1}{L}} u_2^{\frac{1}{L-1}} \ldots U_k^{\frac{1}{L-k+1}}. \label{stepup}
\end{split}
\end{equation}
Further optimization of the simulation algorithm is illustrative because it provides intuition for theoretical derivation of the RTP distribution. This optimization is achieved by using the Markov property of order statistics. Specifically, the unordered set $\{P_{1},\Nb P_{2},\Nb\ldots,\Nb P_{k}\Nb\mid\Nb P_{(k+1)}=p_{(k+1)}\}$ behaves as a sample of $k$ independent variables, identically distributed as $\text{Uniform}\left(0, p_{(k+1)}\right)$. This is a usual step in analytic derivations of product truncated distributions, and it follows by averaging over the density of $P_{(k+1)}$
(this approach was employed earlier by Dudbridge and Koeleman\cite{dudbridge2003} and by Zaykin et al.\cite{Zaykin2002}). After re-scaling,
\begin{eqnarray}
  \left\{\frac{P_1}{p_{(k+1)}}, \frac{P_2}{p_{(k+1)}}, \ldots, \frac{P_{k}}{p_{(k+1)}}\right\} \sim \text{Unif}(0,1).
 \label{markov1}
\end{eqnarray}
The capital $P_i$ notation is used here to emphasize the fact that the variable is random, while the lowercase $p_{(k+1)}$ refers to a realized value of a random variable, $P_{(k+1)}=p_{(k+1)}$.  Next, given that $P_{(k+1)} \sim \text{Beta}(k+1, L-k)$, minus log of the product of independent conditional uniform random variables will follow a gamma distribution. Specifically,
$$
- \ln \prod_{i=1}^{k} \frac{P_i}{p_{(k+1)}} = k  \ln p_{(k+1)} -\sum_{i=1}^{k} \ln {P_i}, $$
and treating $p_{(k+1)}$ as a constant, $$- \ln  \prod_{i=1}^{k} \frac{P_i}{p_{(k+1)}} \sim \frac{1}{2}\chi_{2k}^2 = \text{Gamma}(k,1). $$ 
The above manipulations reduce the set of $k$ random variables to a set of just two variables: a gamma and a beta. Therefore, the combined RTP \Pv{} can be evaluated numerically by simulating only pairs of beta- and gamma-distributed random variables as follows. We note that
$$-\ln \left(\prod_{i=1}^{k} P_{(i)} \right) = -\ln \left(\prod_{i=1}^{k} \frac{P_{i}}{p_{(k+1)}} \right) - k \ln p_{(k+1)},$$
and define
\begin{eqnarray}
 X &=& P_{(k+1)} \sim \text{Beta}(k+1,L-k) \label{eq:X} \\ 
 Y \mid X &=& -\ln \left(\prod_{i=1}^{k} \frac{P_{i}}{p_{(k+1)}} \right) \sim \text{Gamma}(k,1).  \label{eq:Y}
\end{eqnarray}
The empirical distribution of the product of $k$ values under $H_0$ can then be obtained by repeatedly simulating $X$ and $Y$, and comparing the observed value of $-\ln(w_k)$ to $Z\Nb=\Nb Y\Nb-\Nb k\Nb\ln(X)$ in every simulation. $P_{\text{RTP}}$ would then be defined as the proportion of times simulated values of $Z$ were larger than $-\ln(w_k)$. Surprisingly, one can simultaneously evaluate probabilities for two consequtive partial products, 
\begin{eqnarray*}
   && \Pr(W_{k} \le w), \quad \text{and} \\
   && \Pr(W_{k+1} \le w),
\end{eqnarray*}
by reusing the same pair of random numbers, which follows from the fact that
\begin{eqnarray}
   -\ln \left(\prod_{i=1}^{k+1} P_{(i)} \right) = -\ln \left(\prod_{i=1}^{k} \frac{P_{i}}{p_{(k+1)}} \right) - (k+1) \ln p_{(k+1)}.
\end{eqnarray}
In the latter case, $-\ln(w)$ is compared to $Z = Y-(k+1)\ln(X)$. This simulation method is very fast and approaches the exact solution as the number of simulated pairs increases. Moreover, through these simulation experiments it becomes clear that once one conditions on the observed value of $p_{(k+1)}$, the test statistic is formed as a product/sum of independent random variables. Specifically, Gamma distribution for the $Y$ variable in Eq. (\ref{eq:Y}) appears to be conditional on the observed $X = p_{(k+1)}$ when the pairs $(X, Y)$ are simulated. Alternatively, one can first simulate $X = p_{(k+1)}$ and then generate a test statistic using $k$ uniform random variables, $U_1, U_2, \ldots, U_k$, on (0, $p_{(k+1)}$) interval.

We just described a way to evaluate the RTP distribution by repeated sampling of two random variables to elucidate the idea that the combined RTP \Pv{} can be obtained by integrating out the random upper bound $P_{(k+1)}$ over its probability density function. Random $P_{(k+1)}$ has to be at least as large as $p_{(k)}$ but smaller than one, $p_{(k)} \le P_{(k+1)} \le 1$. After re-expressing $p_{(k)}$ in terms of the observed product $w = \prod_{i=1}^{k} p_{(i)}$, it becomes evident that $w^{1/k} \le P_{(k+1)} \le 1$ because the product is maximized if $p_{(i)} = p_{(k)}$ for all $i = 1, \ldots, k$, so the observed $p_{(k)}$ can be at most $w^{1/k}$. Now, integrating over the beta density, $f(\cdot)$ with parameters $k+1, L-k$, of a single variable $P_{(k+1)}$, we will treat $w$ as a constant:
\begin{eqnarray}
  \Pr(W_{k} \le w) &=& 1-\int_{w^{1/k}} ^1 G_{k}\left\{\ln \left(\frac{t^{k}}{w}\right)\right\} f(t) d t.\label{eq:pk1}
\end{eqnarray}
Next, following a transformation, we can express the integral as an expectation and make the integration limits to be 0 to 1, and thus, independent of $k$:
\begin{eqnarray}
  P_{\text{RTP}}(k) = \Pr(W_{k} \le w) = 1-\int_0^1 G_{k}\left\{\ln \left(\frac{\left[B^{-1}_{k+1}(u)\right]^{k}}{w}\right)\right\} du,\label{eq:wkS}
\end{eqnarray}
where $B^{-1}_{k+1}(\cdot)$ is inverse CDF of $\text{Beta}(k+1, L-k)$ distribution, and $G_k(\cdot)$ is CDF of Gamma$(k,1)$. $P_{\text{RTP}}(k)$ is the combined RTP \Pv{}. Similarly, and following the above note that two partial products can be evaluated at the same time,
\begin{eqnarray}
\Pr(W_{k+1} \le w) &=& 1-\int_0^1 G_{k}\left\{\ln \left(\frac{\left[B^{-1}_{k+1}(u)\right]^{k+1}}{w}\right)\right\} du.\label{eq:wkS1}
\end{eqnarray}
We have now derived simple expressions that involve only a single integral where the integration limits (Eq. (\ref{eq:wkS})) no longer involve a product value $w$ and are conveniently bounded within zero to one interval. Eq. (\ref{eq:wkS}) illustrates that the RTP distribution can be viewed as the expectation of a function of a uniform random variable, $U \sim$Uniform(0,1). If we let $H(u \mid k, w) =G_k\left(\ln \left(\frac{\left[B^{-1}_{k+1}(u)\right]^{k}}{w}\right)\right) $, the unconditional distribution of $W_k$ is
$$\Pr(W_k \le w) = 1 - \int_{0}^{1} H(u \mid k, w)du =1 - E\left\{H(U \mid k, w)\right\}.$$
Therefore, to evaluate $P_{\text{RTP}}(k)$ numerically, one can simply sample a large number of uniform random numbers, $U$, apply the function $1-H(U)$ and then take the mean. The corresponding R code using one million random numbers is:

\begin{singlespacing}
\begin{small}
\begin{verbatim}
mean(1-pgamma(log(qbeta(runif(1e6),k+1,L-k))*k+z,k))
\end{verbatim}
\end{small}
\end{singlespacing}
where \texttt{z} = $-\ln(w)$. Using the integration explicitly, the R code is:
\begin{singlespacing}
\begin{small}
\begin{verbatim}
integrate(function(x,w,k,L) 1-pgamma(log(qbeta(x,k+1,L-k))*k+w,k),0,1,z,k,L)$va.
\end{verbatim}
\end{small}
\end{singlespacing}

\subsection{R code example for computation of  $\text{ART}$ and $P_{\text{RTP}}$}
In the code below, $\text{ART}$ and $P_{\text{RTP}}$ are computed for a vector of six \Pvs{} with $k$=4, \texttt{lW}$ =\sum_{i=1}^{k-1} \ln(P_{(i)})$ and \texttt{Pk}$ = P_{(k)}$:

\begin{singlespacing}
\begin{small}
\begin{verbatim}
Art <- function(lW, Pk, k, L) {
   d = (k-1)*(digamma(L+1) - digamma(k))
   ak = (k-1)*log(Pk) - lW + qgamma(1-pbeta(Pk, k, L-k+1), shape=d)
   1 - pgamma(ak, shape=k+d-1)
}
P = sort(c(0.7, 0.07, 0.15, 0.12, 0.08, 0.09))
L = length(P)
k = 4
Z = sum(-log(P[1:k]))
lW = sum(log(P[1:(k-1)]))
P.rtp = integrate(function(x,y,m,n) 1-pgamma(log(qbeta(x,m+1,n-m))*m+y,m),0,1,Z,k,L)$va
P.ak = Art(lW, P[k], k, L)
\end{verbatim}
\end{small}
\end{singlespacing}
The resulting combined \Pvs{} are $\text{ART}$=0.045 and $P_{\text{RTP}}$=0.047. Note that all six original \Pvs{} are larger than the combined ART and RTP. This example demonstrates that weak signals can form a much stronger one after they are combined.

\subsection{Derivation of the ART-A distribution}
As we discussed, ordered \Pvs{} can be represented as functions of the same number of independent uniform random variables (Eq. \ref{stepup}). This reveals that the $j$th value, $p_{(j)}$, is a function of all $p_{(i<j)}$ and that in a given set of $k$ variables (i.e., conditionally) all information is contained in $k$ independent random variables, $U_1, U_2, \ldots, U_k$. These independent components can be extracted and utilized. Specifically, by using the conditional distribution of $W_i$, which only depends on the two preceding partial products, $W_{i-1}$ and $W_{i-2}$, we define independent variables $Z_i$'s as $Z_i\Nb =\Nb \Pr(W_i\Nb>\Nb w_i\Nb \mid \Nb W_{i-1}\Nb,\Nb W_{i-2}\Nb)$. Successive partial products relate to one another as:
$$ W_k = W_{k-1} -  W_{k-1}\left(1- \frac{ W_{k-1}}{ W_{k-2}}\right)U_k^{\frac{1}{L-k+1}}.$$ 
Since $U^{\frac{1}{L-k+1}} \sim \text{Beta}(L-k+1, 1)$, the conditional density and the CDF for the product is \\
\begin{equation*} 
\begin{split}
f(W_k = x \mid W_{k-1} = t_{k-1}, W_{k-2} = t_{k-2}) & = \frac{(t_{k-1}-x)^{L-k}}{B(L-k+1,1)\left(t_{k-1}(1-
	\frac{t_{k-1}}{t_{k-2}})\right)^{L-k+1}}. 
\end{split}
\end{equation*}
Let
\begin{equation*} 
\begin{split}
1 - Z_i & = \Pr\left(W_i < w_i \mid W_{k-1} = t_{k-1}, W_{k-2} = t_{k-2}\right) \\
& = \Pr\left(t_{i-1} -  t_{i-1}\left(1- \frac{ t_{i-1}}{ t_{i-2}}\right)U_i^{\frac{1}{L-i+1}} < w_i  \mid W_{i-1} =t_{i-1} , W_{i-2} = t_{i-2}\right)  \\
& = \Pr\left(-\ln U_i < -(L-i+1)\ln \left(\frac{1-p_{(i)}}{1-p_{(i-1)}}\right)\right).
\end{split}
\end{equation*}
Then,
\begin{eqnarray*} 
\Pr(W_k \le x \mid W_{k-1} = t_{k-1}, W_{k-2} = t_{k-2}) & = & \int_{t_{k-1}^2/{t_{k-2}}}^x f(W_k = x \mid W_{k-1} = t_{k-1}, W_{k-2}) dx \\
& = & \frac{1}{B(L-k+1,1)\left(t_{k-1}\left(1-\frac{t_{k-1}}{t_{k-2}}\right)\right)} \\
& \times & \int_{t_{k-1}^2/{t_{k-2}}}^x (t_{k-1}-x)^{L-k} dx \\
& = & 1 - \left(\frac{t_{k-1} - x}{t_{k-1}\left(1-\frac{t_{k-1}}{t_{k-2}}\right)}\right)^{L-k+1} \\
&= & 1 - \left(\frac{1-p_{(k)}}{1 - p_{(k-1)}}\right)^{L-k+1}.
\end{eqnarray*}
We now obtained a transformation to a new set of independent uniform ($0-1$) random variables. 
$$Z_i = \left(\frac{1-p_{(i)}}{1 - p_{(i-1)}}\right)^{L-i+1},$$ with
$$Z_1 = \left(1 - p_{(1)}\right)^L.$$
Next, define $Y = \sum_{i=1}^k G_{\lambda_i}^{-1} (1 - Z_i)$, where  $G_{\lambda_i}^{-1}$ is the inverse gamma CDF with the shape ${\lambda_i}$ and the scale 1. Under $H_0$, $Y$ has a gamma distribution with the shape equal to the $\sum_{i=1}^k \lambda_i$. The combined $P$-value is now obtained as:
\begin{eqnarray}
   1-G_{\sum_{i=1}^k \lambda_i}\left(\sum_{i=1}^k G_{\lambda_i}^{-1} \left(1 - Z_i \right)\right). \label{gammamethod}
\end{eqnarray}
When $\lambda_i$ is large, the gamma CDF approaches the standard normal CDF, which motivates the inverse normal transformation. The quantiles will be calculated by using $\lambda_i \Phi^{-1} (1 - Z_i)$, as an approximation to $G_{\lambda_i}^{-1}(1 - Z_i)$ for large $k$. The inverse normal method is useful for the reason that the joint distribution of the partial sums can be derived in a standard way to evaluate the adaptive ART (ART-A) \Pv{}. For the ART-A, we define partial sums as:  
$$S_k = \sum_{i=1}^k \lambda_i \Phi^{-1} (1 - Z_i),$$
where $\Phi^{-1}(\cdot)$ is inverse CDF of the standard normal distribution. Then, under the null hypothesis, $\matr{S}=(S_1,S_2,...,S_k)^T$ follows a multiviate normal distribution MVN$(\mathbf{0}, \matr{\Sigma})$, with $\matr{\Sigma} = \matr{FWF}^{T}$ and \\
$$\matr{F} = \begin{bmatrix}
1 & 0 & \cdots & 0 &0 \\
1 & 1 & \cdots & 0 &0 \\
\vdots & \vdots & \ddots &\vdots & \vdots \\
1 & 1 & \cdots & 1 & 0 \\
1 & 1 & \cdots & 1 & 1 
\end{bmatrix}
\text{,\,\,\,\,}
\text{diag}(\matr{W}) =\begin{bmatrix}
\lambda_1^2 &  & &   \\
& \lambda_2^2 &  &   \\
&  & \ddots &   \\
&  &    & \lambda_k^2 
\end{bmatrix},$$ where $\lambda$ are weights. In our simulation experiments, we set all $\lambda_i = 1$, however one may take advantage of some information about the effect size distribution, if that is available. If power is high, but the signal is sparse, it would be expected that true signals may tend to rank among the smallest \Pvs{}. In this case, one possible sequence of weights is $\lambda^2_{k-i+1} = \frac{k}{k-i+1}$.  Such weights that emphasize partial sums with few terms can also be used in certain situations where \Pv{} distribution is expected to be skewed from the uniform (e.g., due to discreteness of a test statistic), with many \Pvs{} being close to one. Finally, the vector $\matr{S}$ can be standardized as $T_i = S_i / \sigma_i$, where $\sigma_i$ are the diagonal elements of $\matr{\Sigma}$, then $\matr{T} \sim$ MVN$(\matr{0}, \matr{R})$, $R_{ij} = \frac{\Sigma_{ij}}{\sqrt{\Sigma_{ii}\Sigma_{jj}}}$. The null distribution of $\matr{T}$ is used to evaluate ART-A by using $\Pr(S_i / \sigma_i > s_i)$ probabilities and to obtain quantiles (significance thresholds) using commonly available MVN distribution functions (e.g., \texttt{mvtnorm} R package).\cite{mi2009mvtnorm}

\subsection{Simulation setup}
We performed $B$=100,000 simulations to evaluate the Type I error rate and power of the proposed methods. To study performance of combination methods for independent \Pvs{}, in each simulation, we generated $L$ normally distributed statistics, $X \sim N(\mu, 1)$. The squared values of $X$ follow the chi-square distribution with one degree of freedom and noncentrality parameter $\mu^2$, $X^2 \sim \chi^2_{(1, \mu^2)}$. \Pvs{} were obtained as one minus the CDF of the noncentral chi-square evaluated at $X^2$, or as $P = 2 - \Phi\left(|X| + |\mu|\right) - \Phi\left(|X| - |\mu|\right)$ in terms of the normal CDF. \Pvs{} generated from normal statistics (without squaring them) were also considered, but these results are omitted for brevity, because the resulting ranking of the methods by power was found to be similar. Under $H_0$, $L$ \Pvs{} were sampled from the uniform (0, 1) distribution, which is equivalent to setting $\mu$ to zero. 

To study non-independent \Pvs{}, we simulated $L$ statistics from a mutivariate normal distribution $\text{MVN}\left(\boldsymbol{\mu}, \matr{\Sigma}\right)$ and decorrelated them by eigendecomposition as described in ``Methods'' section. In each simulation, a correlation matrix $\matr{\Sigma}$ was generated randomly by perturbing an equicorrelated matrix $\matr{D}$. Specifically, we added perturbation to equicorrelated matrix $\matr{D}$ with off-diagonal elements $\rho=0.5$ as:
\begin{eqnarray}
   \matr{R} = \matr{D} + \matr{u}\matr{u}^{T},
\end{eqnarray}
where $\mathbf{u}$ is random vector.\citep{bartlett1951} Then, $\mathbf{R}$ was converted to a correlation matrix $\matr{\Sigma}$ with off-diagonal elements $\rho_{ij} = \frac{R_{ij}}{\sqrt{R_{ii}R_{jj}}} = \frac{\rho + u_iu_j}{\sqrt{1+u_i^2}\sqrt{1+u_j^2}}$. The amount of ``jiggle'' in $\mathbf{R}$ depends on the variability of elements in $\matr{u}$. If elements of $\mathbf{u}$ are generated in the range between $-\delta$ and $\delta$, the value of $\delta$ would represent the upper bound for the amount of jiggle allowed between pairwise correlations in $\mathbf{\Sigma}$. In our simulations, we set $\delta=1$, allowing for a mix of positive and negative values of $\rho_{ij}$ in $\matr{\Sigma}$.

In addition, we evaluated power of the methods by using correlation due to linkage disequilibrium (LD) in real data. The 11 $\times $11 correlation matrix was estimated from previously reported haplotype frequencies of eleven SNPs in the $\mu$-opioid receptor (\textit{MOR}) gene.\cite{shabalina2008expansion,kuo2014discovering}. The pairwise LD correlations within \textit{MOR} were generally high and ranged from -0.82 to 0.99. In this set of simulations, we used effect sizes sampled uniformly in the interval from -0.5 to 0.2.

The Type I error rate and power performance were computed based on two $B \times k$ matrices of \Pvs{}, $\mathbf{P}_0$ and $\mathbf{P}_A$, every row of which contained $k$ smallest sorted \Pvs{} out of $L$ tests across $B$ simulations ($L-k$ \Pvs{} were discarded). $\mathbf{P}_0$ stored simulated \Pvs{} under $H_0$ and $\mathbf{P}_A$ under the alternative hypothesis, $H_A$. Taking the product of \Pvs{} in each row, we obtain two $B \times 1$ vectors, $\mathbf{w}_0$, $\mathbf{w}_A$. RTP \Pvs{} were computed based on the empirical CDF (eCDF) of $\mathbf{w}_0$ evaluated at $B$ values of $\mathbf{w}_A$. Power was calculated as the proportion of \Pvs{} that were smaller than the significance threshold, $\alpha$.

Finally, when various combined \Pv{} methods are being compared, it is meaningful to gauge their performance against methods designed for multiple testing adjustments. This is especially relevant with methods that employ truncation due to their emphasis on small \Pvs{}. Therefore, we included the Simes method\cite{simes1986} in our power comparisons because it can be viewed as a combined \Pv{} method. The Simes method tests the overall $H_0$ without a reference to individual \Pvs{}: the $H_0$ is rejected at $\alpha$ level if $P_{(i)} \le i \alpha/L$ for at least one $i$. Equivalently, the overall (or the ``combined'') Simes \Pv{} can be obtained as $\min\{k p_{(i)}/i\}$. The Simes test is a useful benchmark, because it is related to the combined \Pv{} methods with truncation, as well as to multiple testing adjustment procedures. At the extreme, the RTP with $k=1$ becomes equivalent to \v{S}id\'{a}k correction.\cite{sidak1967} 
\v{S}id\'{a}k correction is approximately the Bonferroni correction,\cite{bonferroni1935calcolo} 
for small \Pvs{} and large $L$. The Simes \Pv{} is at least as small as Bonferroni-corrected \Pv{}. In addition, there is a connection of the Simes test to the Benjamini \& Hochberg false discovery rate (FDR)\cite{benjamini1995cfd}, i.e., the Simes test is algebraically the same procedure as the Benjamini \& Hochberg FDR, although the interpretation is different: FDR method determines the largest $i$, such that  $P_{(i)} \le i \alpha/L$, and rejects $H_0$ for all $P_{(j)}$, $j \le i$, to control the expectation of FDR.

\end{document}